\documentclass[10pt]{article}

\usepackage{amsmath}
\usepackage{amssymb}

\usepackage{graphicx}

\usepackage{cite}

\usepackage{color} 

\usepackage[labelfont=bf,labelsep=period,justification=raggedright]{caption}

\makeatletter
\renewcommand{\@biblabel}[1]{\quad#1.}
\makeatother

\date{}

\newcommand{\W}{\mathrm{w}}   
\newcommand{\M}{\mathrm{m}}

\newcommand{\D}{\mathrm{d}}
\newcommand{\E}{\mathrm{e}}

\begin{document}

\begin{flushleft}
{\Large
\textbf{The rate of beneficial mutations surfing on the wave of a range expansion}
}
\\
R\'{e}mi Lehe$^{1,2}$, 
Oskar Hallatschek$^{3}$, 
Luca Peliti$^{1,4\ast}$
\\
\bf{1} Physico-Chimie Curie, UMR 168, Institut Curie, 26, rue d'Ulm,  F--75005 Paris (France)
\\
\bf{2} \'Ecole Normale Sup\'erieure, 24, rue Lhomond, F--75231 Paris Cedex 05 (France)
\\
\bf{3} Biophysics and Evolutionary Dynamics Group,
Max Planck Institute for Dynamics and Self-Organization,
Bunsenstra\ss e 10, D--37073 G\"ottingen (Germany)
\\
\bf{4} Dipartimento di Scienze Fisiche and Sezione INFN, Universit\`a ``Federico~II'', Complesso Monte S. Angelo, I--80126 Napoli (Italy)
\\
$\ast$ E-mail: peliti@na.infn.it
\end{flushleft}

\section*{Abstract}
Many theoretical and experimental studies suggest that range expansions can have severe consequences for the gene pool of the expanding population. Due to strongly enhanced genetic drift at the advancing frontier, neutral and weakly deleterious mutations can reach large frequencies in the newly colonized regions, as if they were \textit{surfing} the front of the range expansion. These findings raise the question of how frequently beneficial mutations successfully surf at shifting range margins, thereby promoting adaptation towards a range-expansion phenotype. Here, we use individual-based simulations to study the surfing statistics of recurrent beneficial mutations on wave-like range expansions in linear habitats. We show that the \textit{rate} of surfing depends on two strongly antagonistic factors, the probability of surfing given the spatial location of a novel mutation and the rate of occurrence of mutations at that location. The surfing probability strongly increases towards the tip of the wave. Novel mutations are unlikely to surf unless they enjoy a spatial head start compared to the bulk of the population. The needed head start is shown to be proportional to the inverse fitness of the mutant type, and only weakly dependent on the carrying capacity. The precise location dependence of surfing probabilities is derived from the non-extinction probability of a branching process within a moving field of growth rates. The second factor is the mutation occurrence which strongly decreases towards the tip of the wave. Thus, most successful mutations arise at an intermediate position in the front of the wave. We present an analytic theory for the tradeoff between these factors that allows to predict how frequently substitutions by beneficial mutations occur at invasion fronts. We find that small amounts of genetic drift increase the fixation rate of beneficial mutations at the advancing front, and thus could be important for adaptation during species invasions.
\section*{Author Summary}
When a life form expands its range, the individuals close to the expanding front are more likely to dominate the gene pool of the newly colonized territory. This leads to the sweeping of pioneer genes across the newly colonized, a process which has been termed gene surfing. We investigate how this effect interferes with natural selection, by evaluating the probability that an advantageous mutant, appearing close to the edge of an advancing population wave, is eventually able to dominate the population range expansion. By numerical simulations and heuristic analysis, we find that the surfing of even strongly beneficial mutations requires that are introduced with a certain spatial head start compared to the bulk of the population. However, as one moves ahead of the wave, one finds fewer individuals which can possibly mutate. As a consequence, successful mutations are most likely to arise at an intermediate position in front of the wave. For small selective advantage, the success probability is enhanced by an even small amount of genetic drift. This effect could be important in aiding adaptation to local conditions in a range-expansion process.
\section*{Introduction}
\label{sec:intro}
While theoretical population genetics has traditionally focused on stable populations, it is increasingly recognized that departures from demographic equilibrium are a source of major changes in the gene pool of natural populations~\cite{Excoffier:2008hw}. Understanding these non-equilibrium scenarios often requires the development of new theoretical models that are beyond the standard methods of population genetics. 

One of the most frequently observed non-equilibrium scenarios are range expansions, which can be triggered by environmental changes, such as the recent global warming, by adaptive sweeps or species invasions~\cite{excoffier2009genetic}. In the simplest case, range expansion take the form of an expansion wave, as first described theoretically by R. A. Fisher~\cite{fisher}. Range expansions lead to a strong reduction in the genetic diversity of the population because the dynamics is dominated by the few individuals that happen to be at the front of the wave. These \emph{pioneers}~\cite{Hallatschek-theory} are the only ones that have access to empty space and are therefore more likely to proliferate. Moreover, their offspring can, by mere random migration, remain close to the tip of the advancing wave, so that they too can enjoy high resources and continue to proliferate. By this mechanism, the pioneer genotypes are continually transmitted forward and \emph{surf} along with the wave~\cite{edmons,klopfstein}. Thus, all other things being equal, genetic variants which are closer to the front of the wave will have higher probability to fix in the advancing population.

So far, most studies have been concerned with the surfing of neutral alleles. The goal was to understand the patterns created by stochastic drift during range expansion in order to infer past range expansions from observed patterns of genetic diversity. The importance of the surfing of neutral mutations has been quantified by considering the \emph{surfing probability} that a mutation introduced at a specific location close to the front of an advancing population will fix at the front. The larger this probability, the more likely it is that surfing alleles will dominate the gene pool after the range expansion is complete, and the stronger becomes the associated loss in genetic diversity. It was found that surfing is a fast-acting mechanism that is hard to avoid in simple models. Mutation surfing was found to be relevant even in the expansion of large microbial colonies, in which genetic drift would be virtually absent if the population was well-mixed~\cite{hallatschek2007genetic}. In these two-dimensional populations, surfing generates a clear sectoring pattern, later reproduced by simulations~\cite{Excoffier:2008hw,hallatschek2010life}.

It thus seems that the surfing phenomenon can lead to a severe alteration of the gene pool, which may sometimes appear like a ``genetic revolution''~\cite{Excoffier:2008hw}. The fact that new mutations can quickly rise in frequency is phenotypically inconsequential for neutral mutations, but if beneficial or deleterious mutations surf, this can lead to rapid evolution---be it adaptive or maladaptive.  Several simulation studies~\cite{travis, klopfstein,miller2010original} have therefore been carried out to investigate the effect of selection during range expansions. These studies suggest that selection is less efficient at shifting range margins and even deleterious alleles may be able to benefit from the surfing phenomenon. Other simulation studies~\cite{Burton:2008gg,munkemuller:2010dw} have focused in identifying the traits which are more strongly selected close to the expanding front and have suggested that natural selection tends to increase the dispersal and reproduction rates in the expanding population front with neutral or even deleterious consequences for the fitness at carrying capacity (competitive ability). This effect could be enhanced by assortative mating between fast-dispersing individuals~\cite{shine11}. The trade-off between traits selected at the front and those selected in the bulk are consistent with common-garden experiments~\cite{lachmuth}, with observations in the ongoing range expansions of cane toads in Australia~\cite{brown2007invasion} and genealogical records of expanding human populations~\cite{moreau2011deep}. 

These observations raise the question of how fast pioneer populations can potentially adapt at shifting range margins in the presence of recurrent beneficial mutations of a certain effect. This question actually entails two basic theoretical questions that have to be answered jointly. On the one hand, it is necessary to quantify the long term survival of newly introduced beneficial mutations. This leads to an analysis of the surfing probability of beneficial mutations, in the spirit of earlier studies described above. And as previously observed, the surfing probability strongly increases with distance from the bulk the of the wave. The second determinant of adaptation is the mutational input of beneficial mutations, which is proportional to the population density. At shifting range margins, the population density is like the surfing probability strongly location dependent, however, in the opposite direction: The population density decreases approximately exponentially with distance. As a consequence, the surfing \emph{rate} of recurrent beneficial mutations must be controlled by a subtle tradeoff between surfing probability and mutational input. The principal goal of the present study is to establish the first analysis of this tradeoff with the goal to reveal the demographic key parameters that control evolution towards a range-expansion phenotype~\cite{van2010gradual}.

In the first step of our study, we investigate quantitatively the surfing probability of a single mutation arising at a specific position with respect to the front of a population advancing in a linear habitat. We consider just one genetic locus so that recombination does not complicate the dynamics. We concentrate on the fate of mutations that provide an advantage at the front but are neutral in the bulk of the population. As mentioned above, some recent studies have indeed suggest that natural selection during range expansions seems to focus on traits of the pioneer population: e.g., it was shown in \cite{moreau2011deep} that pioneer women in a Canadian range expansion in the 19th century had higher fertility at the front, but not in the bulk. For such front-adjusted mutations, we then evaluate the surfing probability as a function of the position at which the mutation arises and of the linear growth rates $r_{\W}$ and $r_{\M}$ of the wild type and of the mutant respectively. The investigation is carried out by individual-based simulations, augmented by a heuristic mathematical analysis based on branching processes. 

In the second step, we convolute the surfing probability with the density profile of the expanding population waves to predict the substitution rate for beneficial mutations at the front of a range expansion. Ultimately, this substitution rate describes whether the surfing of beneficial mutations is rare or abundant, and thus serves as a proxy for the rate of adaptation during range expansions. 

\subsection*{Model}
\label{sec:model}
Our model is a variant of Kimura's stepping-stone model~\cite{kimura} for a population in a linear habitat, and has been used in Ref.~\cite{Hallatschek-theory} to quantify the surfing of neutral mutations. In this model, colonization sites (which are called ``demes'') are regularly distributed along the $x$ axis. Due to limited resources, each deme can only carry up to $K$ individuals. Individuals have a certain probability to ``hop'' from one deme to a neighboring one. Within one deme, logistic, stochastic growth is assumed. Namely, if $n_{\W}$ is the number of wild~type individuals in a given deme, and $n_{\M}$ the corresponding number of mutants, we define the corresponding ratios by $w=n_{\W}/K$ and $m=n_{\M}/K$. Then the average growth {rates} of wild~types and mutants per unit time are given by $r_{\W}w(1-w-m)$ and $r_{\M}m(1-w-m)$, respectively. This description assumes that the individuals are haploid, but the model describes also diploids, if the fitness of the heterozygote is equal to the mean of the fitness of the homozygotes, and if $K$ is taken to mean the double of the carrying capacity of the deme.

In order to implement this model, we use a discrete algorithm, which is similar to that used by Hallatschek and~Nelson~\cite{Hallatschek-theory}. We consider a \emph{box} made up by $M$ neighboring demes, and kept centered on the advancing population wave as explained below. Each deme is filled with $K$ \emph{particles}, which can be of three types: wildtype, mutant and vacancies. (The presence of vacancies means that the deme is not yet saturated and that the population can still grow within it.) Then the state of the box is updated at each time step according to the following process. 
\begin{description}
     \item[Migration step:]  Two neighboring demes are chosen at random. Within each of these demes, a particle is chosen at random, then those two particles are exchanged. (If the two particles chosen are of the same type, this leads to no change.)
     \item[Duplication step:] One deme is chosen at random. Within this deme, two particles are chosen at random, then the second particle is replaced by a duplicate of the first one, with probability $p$. (Again, if the two particles chosen are of the same type, this leads to no change.) The probability $p$ is equal to one for all processes, except for the replacement of a wild~type or a mutant by a vacancy, which happen with probability $p=1-r_{\W}$ and $p=1-r_{\M}$ respectively. It is {possible} to show that this choice indeed results in average growth rates of the form $r_{\W}(1-w-m)$ for wildtype and $r_{\M}(1-w-m)$ for mutants.
\end{description}  
Notice that the probability that a mutant replaces a wildtype individual is equal to that of the opposite event. Therefore, in a full deme, mutants have no competitive advantage over wildtype individuals. However, the relative proportion of mutant and wildtype individuals will be subject to stochastic fluctuations. We define our unit of time so that the diffusion constant of the particles is equal to one.
\section*{Results}
\label{sec:results}
Let us consider a single mutant introduced at a position $x$ measured from the advancing population wave at a certain time. Our first goal is to evaluate the probability $u(x)$ that this mutant becomes the ancestor of the population in the front of the wave. The function $u(x)$ thus represents the probability that the mutation becomes fixed in the front population, given it arose at location $x$.  One can consider this \emph{surfing probability} as a spatial analog of the classical Haldane formula for the fixation probabity of mutant genotypes  in well-mixed populations~\cite{haldane1927mathematical}. Relying on 1D simulations, Hallatschek and Nelson~\cite{Hallatschek-theory} searched for an analytical expression of this specific function. Their work focused on the relatively simpler case of neutral mutations. Progress even in this neutral case was only possible through approximations~\cite{Hallatschek-theory}. However, most recently an exact approach could be devised that relies on modifying the logistic interaction in the population dynamics~\cite{Hallatschek-linear}.

\begin{figure}[hbt]
\begin{center}
     \includegraphics{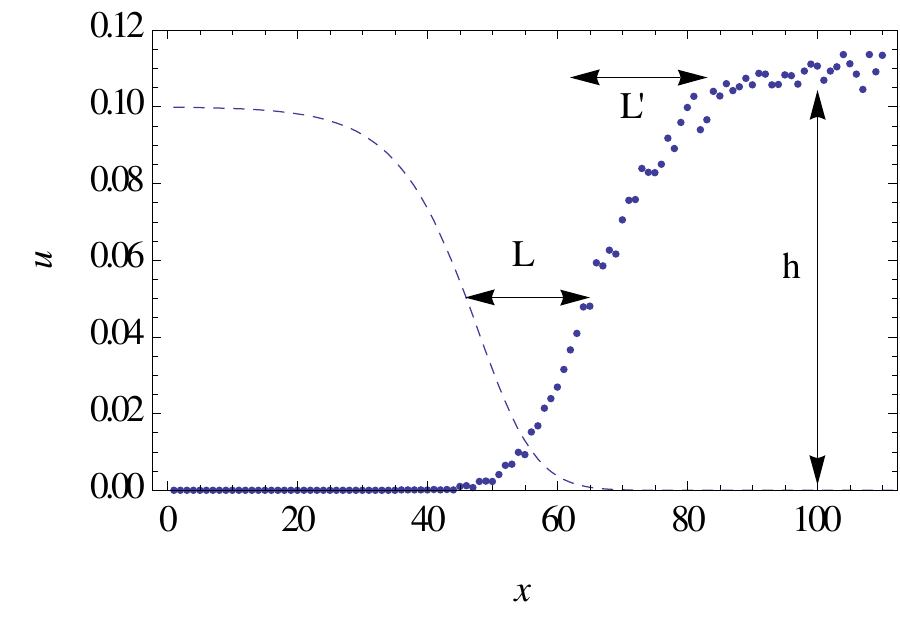}  
\end{center}
\caption{{\bf Fixation probability $u(x)$ as a function of the position $x$ where the mutant is introduced (for $r_{\W} = 0.1$, $r_{\M} = 0.11$ and $K=100$).} The wildtype wave profile (arbitrarily rescaled) is shown by the dashed line. The probability profile $u(x)$ virtually vanishes in the bulk of the wave, but suddenly rises at a characteristic distance $L$ from the front, and then saturates. The shape of the function $u(x)$ is characterized by $L$, defined as the distance between the two points at which the curves reach half of their saturation values, by the characteristic width $L'$ over which the curve rises, and by its saturation height $h$.}
\label{curve}
\end{figure}

\begin{figure}[htbp]
\begin{center}
  \includegraphics{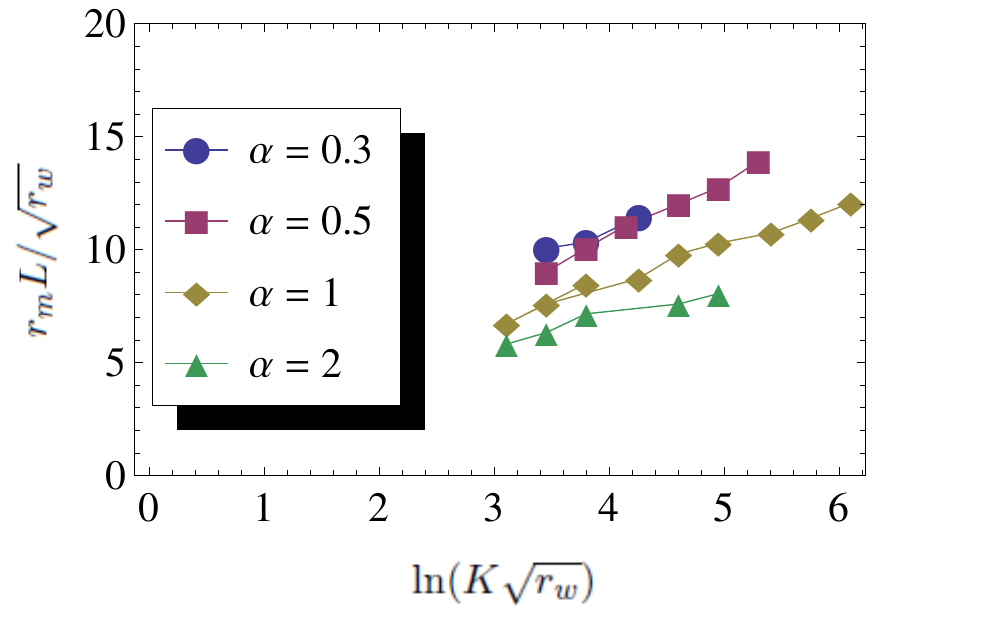} 
\end{center}
\caption{{\bf The measured characteristic distance $L$, rescaled by the factor $r_{\M}/\sqrt{r_{\W}}$, as a function of $\ln(K\sqrt{r_{\W}})$.} The data shown corresponds to various values of $K$, $r_{\W}$ and $r_{\M}$. Values of $K$ range from 100 to 1000, and values of $r_{\W}$ and $r_{\M}$ range from 0.05 to 0.5. The data points corresponding to the same value of $\alpha\equiv r_{\M}/r_{\W}$ group on straight lines, with slopes weakly dependent on $\alpha$. Compare~(\ref{shift-eqn}).}
\label{shift}
\end{figure}

Figure~\ref{curve} shows the typical shape of $u(x)$ for a weakly beneficial mutant, and defines the different characteristic parameters of the curve. It can be seen that, when the mutant starts in the bulk of the wave, the probability that it fixes in the front virtually vanishes. This is indeed what would be expected from the qualitative picture of the surfing mechanism. However, when the mutant starts at a distance of order $L$ from the front, this probability starts to increase. This is due to the fact that, in this case, the mutant population has access to more empty space and is likely to grow for a while, before the wildtype front eventually reaches it. Finally, far ahead of the front, this probability saturates. Importantly, our simulations revealed that the saturation value $h$ of $u$ is always equal to $r_{\M}$. An interpretation of this fact is given in the discussion.
We evaluated the dependence of the characteristic length $L$ as a function of the parameters defining the model. As shown in figure~\ref{shift}, the variations of $L$ are consistent with the expression:
\begin{equation}
\label{shift-eqn}
	L \approx 2\frac{\sqrt{r_{\W}}}{r_{\M}} \ln(K\sqrt{r_{\W}})\, f(\alpha),
\end{equation}
where $f(\alpha)$ is a slowly decreasing function of $\alpha\equiv r_{\M}/r_{\W}$, with $f(1) = 1$. In this expression, $K$ is the carrying capacity of each colonization site (``deme''), and $r_{\W}$ and $r_{\M}$ are, respectively, the wildtype and mutant growth rates (see Methods). A similar expression is derived from a simple model in the discussion. It can be seen that the dependence on $r_{\M}$ is consistent with intuition: the fitter the mutants, the fewer resources they will need in order to invade the front. Therefore, they may start closer to the front, and still have a substantial probability to fix. The dependence on $K$ is less intuitive. However, it was already remarked in ref.~\cite{Hallatschek-theory} that $u(x)$ shifted away from the front for increasing values of $K$. Within the neutral frame work of ref.~\cite{Hallatschek-theory}, an approximate (but general) expression for $u(x)$ in terms of the wave profile was obtained that was quantitatively consistent with these observations.
\subsection*{The basic surfing scenario}
\label{subsec:scenario}
The results just described as well as the direct inspection of particular realizations, guided us in drafting a rough scenario for the fate of the mutants:
\begin{itemize}
\item When the mutants start behind the front (i.e., in the bulk of the wave), it is practically impossible for them to grow, since they are surrounded by nearly full demes. The dynamics is dominated by random birth and death, with no net growth rate, and the population is bound to die out at some point. Therefore the survival probability vanishes.
\item When the mutants start immediately ahead of the front, they have access to partially empty demes and \emph{may} grow for a while. More precisely, during a first stage in which the number of mutants is still low, the population may die out, due to stochastic fluctuations in births and deaths. However in realizations in which that does not happen, the population reaches a number for which fluctuations are negligible, and thus enters a second stage in which it grows rapidly. Yet, in this case, while they grow, the advancing wildtype wave is progressively reaching them. As a result, they will soon be surrounded by full demes, and will not be able to grow anymore. They are therefore left behind. On the whole, the survival probability is relatively large (but still smaller than 1, due to stochastic death in the first stage), while the fixation probability $u$ is small.
\item Eventually, when the mutants start far ahead of the front (at a distance larger than $L$), they have---as before---the possibility to grow rapidly, but it also takes a longer time for the wildtype wave to reach their starting position. Meanwhile, the mutant population may grow to such a large number that they can actually stop the wildtype wave, and develop their own advancing front. In other words, if the mutants survive the stochastic fluctuations of the early stage, they are certain to reach fixation. Therefore the fixation probability $u$ equals the survival {probability} $v$ in this case.
\end{itemize}
We can now begin to provide explanations for the quantitative results of our simulations.

\subsection*{Maximal surfing probability in the wave tip}
\label{subsec:saturation}
According to the basic scenario outlined above, mutants arising far in the tip of the wave fix depending on whether or not they avoid a stochastic death in the first stage of their growth. Notice that the presence of a wildtype wave plays no role here, since it has not yet reached this position. In a large well-mixed population, this survival probability is simply given by $r_\W$ for a branching process with growth rate $r_\W$ and death rate $1$, which is a classical Haldane formula for the establishment probability of a beneficial mutation~\cite{haldane1927mathematical}. This standard result remains unchanged in the present spatial model with local logistic growth, as is shown in the subsection on nondimensionalized equations by a simple argument. Indeed, our simulations show that the probability of survival (and fixation) probability saturates at the value $r_\W$ for sufficiently beneficial mutations in the tip of the population wave.

\subsection*{Onset of surfing in the tip of the wave}
\label{subsec:L}
In the Results section, we defined $L$ as the typical distance, measured from the front, where the surfing probability $u(x)$ changes from 0 to its maximal value $r_{\M}$. In other words, mutants have very small chance to reach fixation if they are introduced at $x <L$, whereas they will almost surely fix if they start at $x>L$, provided that they survive the stochastic fluctuations in the first stage of their growth. In the basic scenario described above, we suggested that this meant in fact that mutants starting further than $L$ have enough time to grow, so that they are then numerous enough to stop the advancing wildtype wave.

This argument can be turned into a quantitative estimate of the magnitude of $L$. According to the classical Fisher wave speed and our numerical measurements, our model (cf.~equation (\ref{ad-system1})) implies that the wildtype wave propagates at a velocity $v \approx 2 \sqrt{r_{\W}}$. Therefore, the wildtype wave will reach the growing mutant population at a time $t_0 = \Delta x_0/v$ (where $\Delta x_0$ is the distance from the front to the starting position). Let us now estimate how much the mutant population will have grown before this arrival time $t_0$ of the wild type wave. To this end, we assume that the mutant clone grows unaffected by the wildtype population up until time $t_0$. Then, the total mutant population $N_{\M}(t)$ grows exponentially on average, according to $\langle N_{\M} \rangle= \exp(r_\M t),$ $t\ll t_0$. However, we know from section \ref{subsec:saturation} that, after some time, the mutant population is non-zero in only a fraction $r_{\M}$ of the realizations. Therefore, $\langle N_{\M} \rangle = (1-r_{\M})\times 0 + r_{\M}\times \langle \bar{N}_\M \rangle$, where $\langle \bar{N}_\M \rangle$ is the average over realizations in which the mutant population has not died out. Thus we have $\langle \bar{N}_\M \rangle=\exp(r_\M t) /r_\M$.

Now we make the simple-minded assumption that the probability to fix is large if the total mutant population has grown above a characteristic number on the order of the typical number $K /\sqrt r_\M$ of mutants in an all-mutant wave before the wildtype wave reaches it (i.e., at time $t_0$) and is small otherwise. Hence, we expect
\begin{equation}
  \label{eq:L-upper-bound}
L \approx 2\frac{\sqrt{r_{\W}}}{r_{\M}} \ln (K\sqrt{r_{\M}})\, f(K\sqrt{r_{\W}},r_{\M}/r_{\W})\;,
\end{equation}
where $f(x,y)$ is a weakly varying function of its two arguments. We will show in the Methods section, that indeed the only relevant parameters that govern the surfing probabilities are $K_{\E}=k\sqrt{r_{\W}}$ and $\alpha=r_{\M}/r_{\W}$, which appear as arguments in \eqref{eq:L-upper-bound}.

Our estimate of the ``edge'' of surfing in \eqref{eq:L-upper-bound} should be considered as an upper bound because a clone may not need to grow to a size as large as the number of individuals in a mutant wave front, as we have assumed in our argument. Nevertheless, the estimate in \eqref{eq:L-upper-bound} sets a useful bound on $L$, which works very well for small populations and large fitness effects, as documented by our data in Fig.~\ref{shift}. 

\subsection*{Functional form of the surfing probability}
\label{subsec:diffeq}
With the onset of surfing in the tip of the wave and the maximum surfing probability $s$, we have discussed two characteristic features of the sigmoidal function $u(x)$. A more detailed analysis is required, however, to describe the transition region where most of the surfing beneficial mutations are generated, which is a pre-requisite for dissecting the substitution rate below. Therefore, we sought for a differential equation that may determine the functional form $u(x)$. An equation of this kind was already found in ref.~\cite{Hallatschek-theory}, for the case of a neutral mutation, on the basis of a backward Fokker-Planck formalism. However, the approach that these authors use is specific to neutral mutations and cannot be extended to the non-neutral case.

For sufficiently beneficial mutations, it is however possible to derive an approximate differential equation for $u(x)$ by employing the theory of branching random walks. To this end, we approximate the proliferation of newly introduced mutant by a linear birth-death process: A mutant at position $x$ has a constant birth rate of $1$ per generation. The death rate on the other hand depends on location. Far in the tip of the wave, the death rate of the mutants approaches a constant of $1-r_\M$, and it approaches $1$ in the bulk of the wave as there is no net growth in the saturated region of the population. By construction of our model, the net $x$-dependent growth rate is given by $r_\M [1-w(x,t)-m(x,t)]$, where $K w(x,t)\equiv n_{\W}(x,t)$ is the number of wildtypes in a deme located at $x$ at time $t$, and $m(x,t)$ is the analogous quantity for the mutants. Thus the net growth rate is in general fluctuating due to the fluctuating occupancy $w(x,t)+m(x,t)$ of deme $x$. We now make two important assumptions. First, we assume that the survival of the mutants is decided early on when the mutant population is so small that we can well approximate its growth rate by the function $r_\M[1-w(x,t)]$, i.e., by neglecting the non-linear effect of the mutant population on its own survival. This approximation is justified when the growth rate advantage of the mutants is sufficiently large, and breaks down in the neutral or nearly neutral case.  Second, we average the growth rates over all realization and assume a growth rate $r_\M[1-\langle w\rangle]$, where $\langle w(x,t)\rangle$ is the average density profile of an all wildtype wave. This simplification holds provided that that the carrying capacity is so large that fluctuations in the wave profile are weak. Under these assumptions, we can use a standard result for branching random walks, namely that the survival probability, which in our case equals the surfing probability $u(x)$, satisfies
\begin{equation}
  \label{eq:BRW-survival-proba}
  0=\partial_x^2u - v_\W\frac{\partial u}{\partial x} +r_\M (1-\langle w\rangle) u - u^2 \;.
\end{equation}
In the Methods section, we provide a heuristic rational of this differential equation, but for a strict derivation the reader is referred to standard text books, such as ref.~\cite{harris2002theory}.

\begin{figure}[!ht]
\centering
\includegraphics{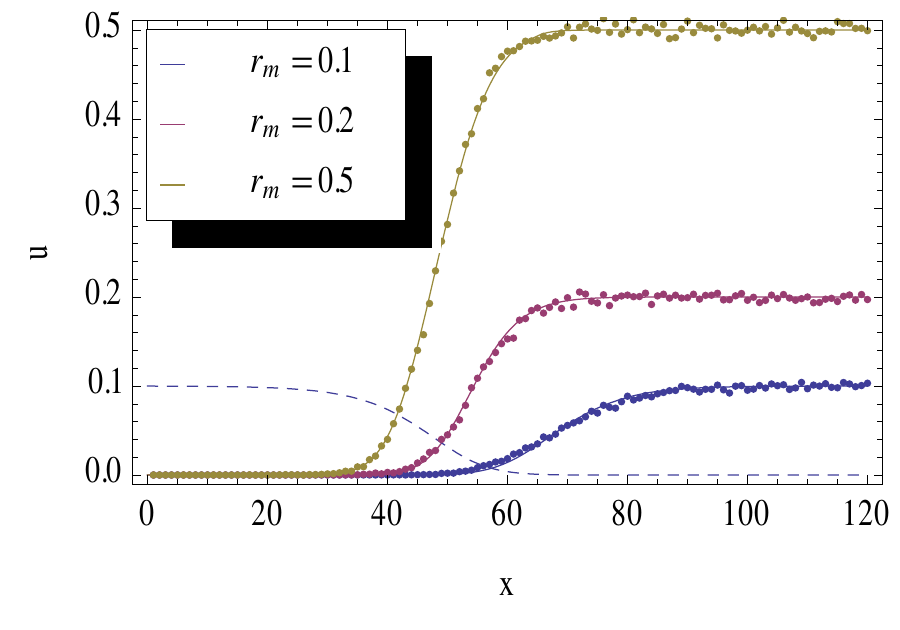}
\caption{{\bf Fixation probability profile $u(x)$ for $K=100$, $r_w = 0.1$, and different values of $r_{\M}$.} The dots represent simulation data and the continuous line corresponds to the result of the numerical integration. The dashed line represents the (arbitrarily rescaled) average profile $\langle w \rangle$ of an all wildtype wave.}
\label{integration-s}
\end{figure}

\begin{figure}[!ht]
\centering
\includegraphics{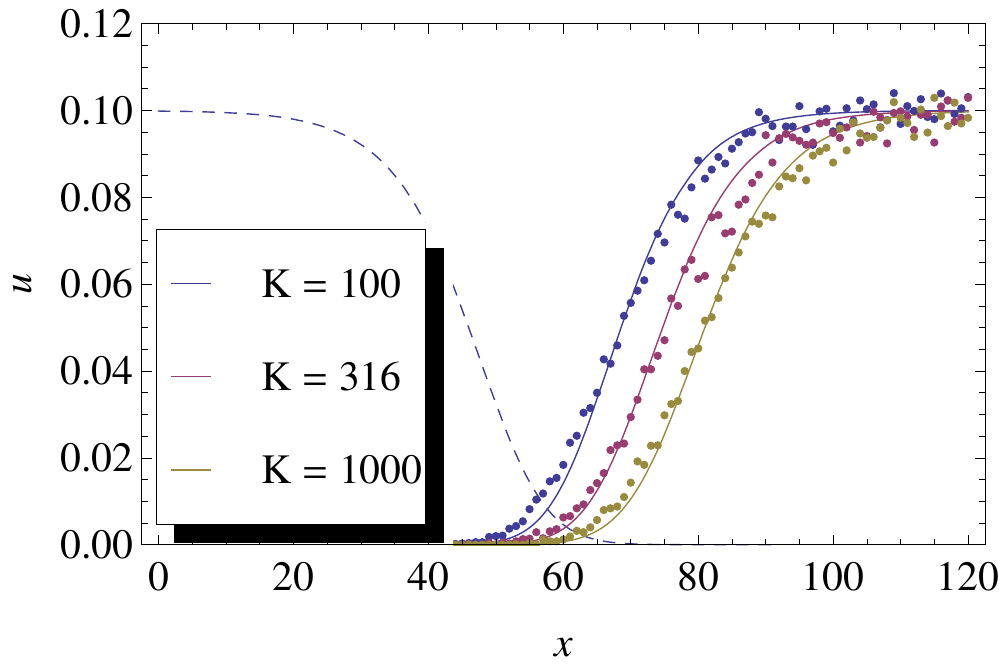}
\caption{{\bf Fixation probability profile $u(x)$ for $r_m = r_w = 0.1$, and different values of $K$.} The dots represent simulation data and the continuous line corresponds to the result of the numerical integration. As before, the dashed line represents $\langle w \rangle_\mathrm{init}$, arbitrarily rescaled.}
\label{integration-N}
\end{figure}

Equation \eqref{eq:BRW-survival-proba} has a form very similar to the differential equation for a deterministic Fisher-Kolmogorov wave running in the $-x$ direction. This explains the overall sigmoidal ``wave profile'' of the function $u(x)$. Notice however that the term $\propto (1-\langle w\rangle)$ approaches $0$ for $x\ll0$ where the wildtype occupancy saturates, $w\to1$. Thus, \eqref{eq:BRW-survival-proba} should be regarded as a classical Fisher-Kolomgorov equation with a \emph{cut-off}~\cite{Derrida}, an observation which will  be important in the following section. To quantitatively compare the branching process theory with our individual based simulations, we integrated equation (\ref{eq:BRW-survival-proba}) numerically. As shown in figure \ref{integration-s} and \ref{integration-N}, the agreement is very good, and remains so when the parameters $K$ and $r_m$ are varied as long as $\alpha>1$.
\subsection*{Rate of substitutions}
\label{sec:rate-substitutions}
As a proxy for the speed of adaptation at shifting range margins, we finally ask how frequently beneficial mutations \emph{fix} in the pioneer population for a given mutation rate $U_\mathrm{b}$. Clearly, the surfing probability $u(x)$ is one important factor as it governs the chances of success for a mutation inserted at location $x$. We have seen that, generically, $u(x)$ steeply increases towards the tip of the wave due to the location advantage appreciated there. However, only few individuals reside in the tip region and can thus provide mutational input for adaptation. This effect is described, of course, by the wave profile $n_{\W}(x)=K w(x)$. The product $u(x)n_{\W}(x)$ describes the tradeoff between the higher success probability in the tip and the higher mutational input in the bulk of the wave. More precisely, the integral
\begin{equation}
  \label{eq:substitution_rate}
  G\equiv \int \D x\; \left<u(x)n_{\W}(x)\right>=K\int\D x\; \left<u(x) w(x)\right> 
\end{equation}
controls the substitution rate $R=U_\mathrm{b} G$ for beneficial mutations of effect $s$ and mutation rate $U_\mathrm{b}$. 

As argued earlier, for sufficiently beneficial mutations, the survival of a beneficial mutation is well-described by our mean-field description that only depends on the mean $\left<w(x)\right>$. We may thus approximate $G$ by setting $\int \D x\,\left<u(x)w(x)\right>\approx \int \D x\,\left<u(x)\right> \left<w(x)\right>$, and use our above results for the average survival probability and population density to estimate the integral on the right hand side. The value of $G$ is plotted in Fig.~\ref{fig:G/K} as a function of the selective advantage of the mutants.  These results show that, for carrying capacities ranging from $K=100$ to $K=1000$, the substitution rate depends only weakly on selection coefficients. Even for selection coefficients of 10\%, the substitution rate in the most dense population ($K=1000$) is merely increased by a factor 4 compared to the neutral base line. Also note that the substitution rate does increase more slowly than linear with population size (as parameterized by $K$) quite in contrast to well-mixed population models (in the absence of clonal interference~\cite{sniegowski2010beneficial}).

Our simulated data for $G$ are hard to model from first principles, as this would require a solution to the long-standing problem of noisy Fisher waves for rather small values of $\ln K$~\cite{Hallatschek-linear}. However, for large carrying capacities such that $\ln K\gg 1$, where genetic drift is weak, an analytical approach is feasible.  The analysis, described in the Methods section, not only allows us to answer the question as to how the substitution rate $G$ behaves in the deterministic limit, or relatively close to it. It also provides us with a qualitative picture of how genetic drift, mutations and selection compete during a population expansion. These asymptotic results are meant to guide the intuition as to how weakly selection affects the subsitution process.
\begin{figure}[!ht]
\centering
\includegraphics[scale=1]{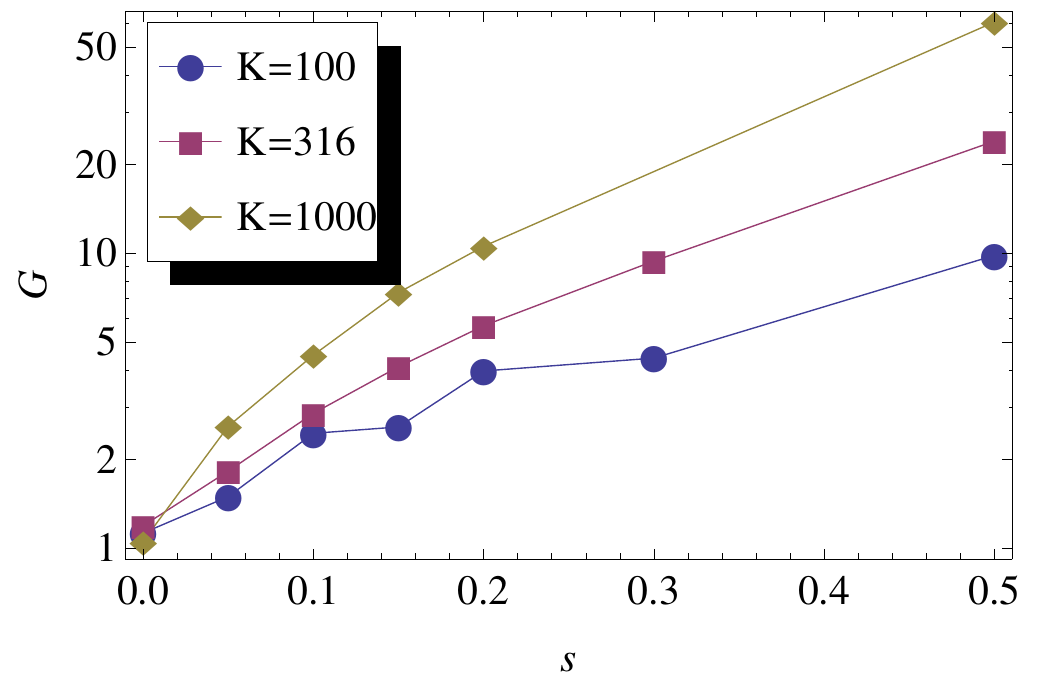}
\caption{{\bf The substitution rate at the front of an advancing population compared to the neutral substitution rate is described by the function $G\equiv K\int\D x\; \left<u(x)\right>\left<w(x)\right>$, which is displayed here as a function of the selective advantage $s$ of the mutants.} Notice that the $G$ axis is logarithmic. Blue, red and yellow symbols correspond to the carrying capacity $K=100$, $K=316$ and $K=1000$. All curves approach $1$ in the neutral case, $s=0$, in which substitution and mutation rate equal. Notice the rather slow increase of substitution rates with increasing selection coefficient, for small values of $s$: even for $s=10\%$ and the highest carrying capacity, the substitution rates are merely 4 times higher than the neutral baseline, illustrating the ineffectiveness of selection at expanding fronts. For larger selection coefficients, however, the substitution rate grows roughly exponentially with $s$. }
\label{fig:G}
\end{figure}
\begin{figure}[!ht]
\begin{center}
	\includegraphics{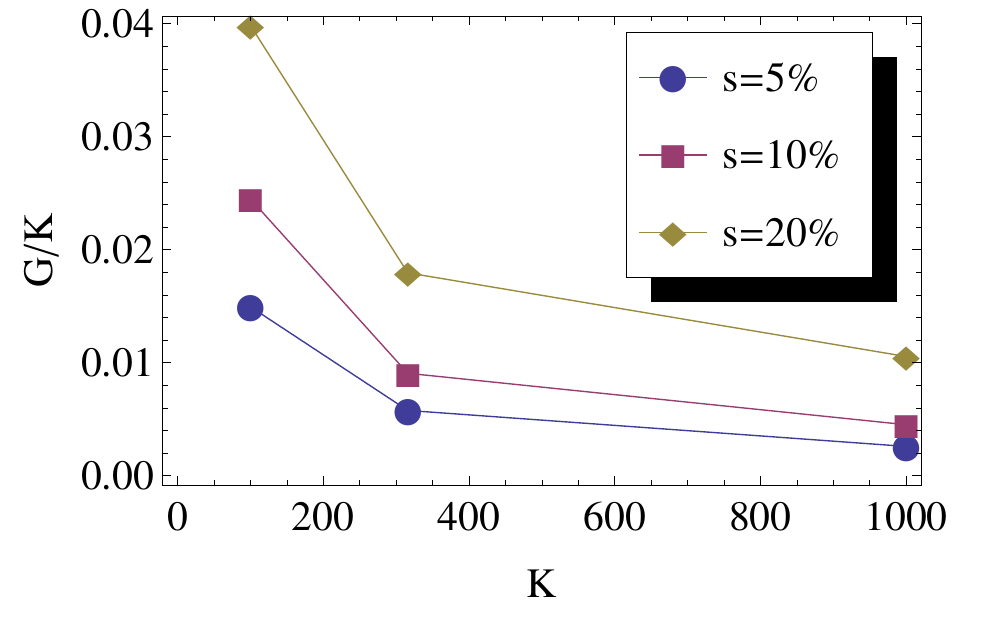}
\caption{{\bf The rate $G$ of surfing events (defined in equation \eqref{eq:substitution_rate}) divided by the deme carrying capacity $K$ as a function of $K$ for several values of the selective advantage $s$ of the mutants.} It appears quite clearly that the substitution rate increases less than linearly with $K$, as suggested by equation~(\ref{eq:G-noisy}).}
\label{fig:G/K}
\end{center}
\end{figure}

\section*{Discussion}
\label{sec:discussion}
When a beneficial mutation arises in the front of an expanding population, it has a high risk of being immediately lost from the front population either by extinction or because the mutant clone cannot keep up with the shifting wave front. Rarely, however, mutants become entirely fixed in the front population, a phenomenon referred to as gene surfing.  In this paper, we have studied the results of a one-dimensional individual-based simulation to measure and explain i) the probability of surfing of a newly introduced beneficial mutations on a population range expansion and ii) the rate of these surfing events if beneficial mutations occur at a certain rate and have a certain effect.

In agreement with earlier studies~\cite{Hallatschek-theory,klopfstein, travis, miller2010original}, we found that the probability of surfing crucially depends on the location of the first mutant with respect to the advancing wave. We have quantified this location advantage in two ways. First, we estimated heuristically the spatial head start required for a clone of beneficial mutations to grow large in the wave tip before the bulk of the wave arrives. This head start was found to be inversely proportional to the growth rate of the mutants and only grows logarithmically with the carrying capacity. If mutations arises sufficiently far ahead of the front of a population-expansion wave, they can fix even if fitness effects are small, which is consistent with earlier observations~\cite{klopfstein, travis, miller2010original}. A more systematic and accurate analysis based on the theory of branching processes could be given to describe how fast the surfing probabilities rise as one moves into the tip of the wave until it eventually saturates. Further analysis, reported in the Methods section, shows that in the deterministic limit of infinite carrying capacities, the characteristic distance at which surfing becomes significant scales as $s^{-1/2}$ for small selective advantage $s$ (cf.~equation \eqref{eq:L-asymptotically-2}). For any reasonable carrying capacity, however, surfing probabilities are found to be significantly higher than expected from a deterministic analysis, which shows that genetic drift is essential for the surfing of weakly beneficial mutations.

Our analytical description of the location-dependent survival probability enabled us to get at our second key question: At what rate do surfing events occur for a given mutation rate and selective advantage? This rate of surfing events may be viewed a proxy for how quickly a population may evolve toward a range expansion phenotype~\cite{van2010gradual}. The surfing rate is determined by two factors. One is, of course, the surfing probability, which increases towards the tip of the wave, the other is the mutational process by which new potential surfers are introduced. Clearly the mutational supply is highest in the bulk of the wave because of its saturated population density, but there the surfing probability is lowest. It turned out that, due the trade-off between both effects, most surfers are generated at an intermediate position within the front of the wave. We were able to determine analytically the substitution rate for large populations and small mutational fitness effects. This analysis shows that, in the deterministic limit, surfing rates for small selection coefficients are strongly suppressed. Mathematically, this is manifest in an essential singularity of the substitution rates at vanishing selection coefficients. For large but finite carrying capacities, however, substitution rates are strongly increased due to even tiny amounts of genetic drift. Our theory predicts a generally quite strong positive correlation between surfing rates and genetic drift (as quantified by inverse carrying capacities) for small selection coefficients. Interestingly, our simulations show that this correlation is qualitatively inverted for large selection coefficients: Very large effect mutations do not require genetic drift to prevail, so that their rate is mainly controlled by the mutational supply which increases with increasing carrying capacities. However, our results suggests for beneficial mutations of intermediate and small effects that long-term survival during a range expansions is mostly a matter of luck to arise far in the wave tip than of fitness.

In summary, we have for the first time analyzed not only the fate of newly introduced mutations, but also the rate of surfing events for a given mutation rate. Our results suggest that genetic drift is not required to promote mutation surfing of strongly beneficial mutations for which selection is strong enough. Importantly, however, our results suggest that some amount of genetic drift strongly increases substitution rates at advancing fronts for weakly beneficial mutations and thus can be important for promoting adaptation towards an invasion phenotype. 

Finally, we discuss the assumptions at the base of our study, and its possible generalizations. First, we only considered mutations that are beneficial to the pioneer population but neutral for the bulk population. Several experimental studies suggest that such mutations towards a range-expansion phenotype are actually disadvantageous in the bulk of the population~\cite{lachmuth,brown2007invasion,moreau2011deep}. While such mutations gradually disappear from the bulk population, we expect that their surfing propensity will be almost identical to mutations that are neutral or beneficial in the bulk. This is because the bulk phenotype matters so far from the wave tip that it cannot influence the genetic composition of the wave tip. The analysis would change qualitatively if the selective advantage in the bulk is so large that the ensuing genetic wave of the beneficial mutation within the saturated bulk population would be faster than the range expansion. However, this situation only occurs for extreme selective differences on the order of one. 

We have also assumed that population expansions proceed according to R.A. Fisher's standard model, in which the Malthusian growth rate of individuals in the tip of the wave is constant. However, many species are characterized by a reduced Malthusian growth rate when densities become too small. This effect arises when individuals need to cooperate with others in order to proliferate, for instance in the case of sexual reproduction. Such Allee effects~\cite{allee1931co} have been found to considerably lessen the role of genetic drift in the gene surfing phenomenon: The \emph{effective} population size associated with the expanding population front was strongly positively correlated with the strength of the Allee effect~\cite{Hallatschek-theory}. We expect that such Allee effects will also alter surfing probability and rates of beneficial mutations, because they lessen the extreme location advantage of mutations arising in the far wave tip. As a consequence, the surfing beneficial mutations arise closer to the bulk of the population for stronger Allee effects. Also the total rate of surfing events would be strongly increased. We thus expect that larger Allee effects will significantly enhance adaptation towards a range expansion phenotype.

Another interesting extension of our study concerns expansion waves in planar habitats. In this case, the location advantage for deleterious mutants is likely to be less relevant, since the wildtype population is able to overcome the mutant and constrain it to a bounded region. As in the one-dimensional case, successful long-term surfing of deleterious mutations will require that the mutant clone takes over the entire colonization front. As a consequence, the surfing probability will sensitively depend on the habitat's extension transverse to the expansion direction. Also, the analysis of the surfing of beneficial mutations will be more complex: Surfing beneficial mutations give rise to \emph{sectors}~\cite{hallatschek2010life} with sector angles that characterize their selective advantage against the surrounding wild type population. Furthermore, at any given time, some parts of the colonization front will be more advanced than others, due to the inevitable random front undulations. If a mutation arises in one of those more advanced region of the habitat, it will have higher long-term surfing probabilities than in the less advanced regions. Nevertheless, simulations of the kind carried out in this study should quite generally allow to investigate the establishment probabilities in any model of expanding populations.

\section*{Methods}
\label{sec:methods}

We initialize our simulations by letting the demes in the left half of the box be full of wildtype individuals, while the demes in the right half of the box are empty (i.e., full of vacancies). Thus, the initial configuration evolves into a wave profile that propagates to the right. The algorithm follows the wave front by shifting the box at the same velocity, by introducing from time to time new empty demes at the right extremity while removing the leftmost demes. In the subsection on nondimensionalized equations, we show that our simulations can be described a set of stochastic differential equations. The form of these equations show that, although our model is characterized by three parameters, $K$, $r_{\W}$ and $r_{\M}$, there are in fact only two control parameters: The relative fitness $\alpha=r_{\M}/r_{\W}$ measures the growth rate advantage of mutants. The parameter combination $K_\E=K\sqrt{r_{\W}}$ quantifies the strength of number fluctuations in the tip of the wave. $K_\E$ is analogous to the parameter ``$N_\E s$'' in many well-mixed population genetic models, with the replacements $s\to r_\W$ and $N_e\to K /\sqrt{r_\W}$. The relevant population size $K/ \sqrt{r_\W}$ represents the typical number of individuals in the nose of a purely wildtype wave, because $K$ is the occupancy of saturated demes and $1/\sqrt{r_{\W}}$ measures the width of the wave front in units of deme sizes. When no mutant is present, the dynamics reduces to the well-known noisy Fisher-Kolmogorov-Petrovski-Piskounov (FKPP) equation~\cite[p.400]{Murray} in one spatial dimension. This is confirmed by control simulations of all wildtype waves, which show that the velocity of the wave tends to $2\sqrt{r_{\W}}$ for large $K$, consistent with the known deterministic wave speed of FKPP waves.
 
To investigate the surfing phenomenon, we studied the wave propagation under the influence of newly introduced mutants. Specifically, a mutant was added in a chosen deme within the co-moving simulation box once the wave has reached a steady profile. The fate of the mutant clone was then recorded. Three types of final events were distinguished:
\begin{description}
\item[Fixation:] the mutants invade the front of the wave and the wildtype population is left behind. A given realization is considered as a fixation event if no wildtype individual remains in the box at the end of the simulation.
\item[Failure:] the mutants survive, but they fail to invade the front and are left behind. This corresponds to realizations in which no mutant remains in the box at the end \emph{but} in which it has been detected that \emph{at least one} mutant has crossed the left boundary of the box, for instance when the box was shifted.
\item[Death:] the mutant population dies out due to stochastic fluctuations. This corresponds to the remaining realizations in which the mutants disappear before reaching the left boundary.
\end{description}
Failure corresponds quite likely to a situation in which the mutant population eventually dies off, due to neutral sampling fluctuations. Even if the mutants had a higher growth rate than the wild types in full demes, if they are able to establish a mutant population in the bulk, they would then expand by Fisher waves with a speed determined by the difference in growth rates with respect to wild types. This wave will not have a chance to catch the wave front unless this difference is unreasonably large. Thus it is safe to neglect the occurrence of failure when focusing on the events on the front of the advancing population wave, also considering that the definition of failure depends on the width of the simulation box. These considerations justify focusing only on the fixation probability at the front of the wave. Therefore, for each starting position $x$, we ran many realizations of the process, and from their results we deduced the probability of fixation. The number of realizations over which the algorithm evaluated those probabilities was usually set to 10\,000, for each position $x$. The position $x$ was then varied to obtain the dependence of this probability on the starting position.

\subsection*{Master equation and its nondimensionalized expression}
\label{subsec:nondimensionalized_equations}

In the present subsection we show that the different parameters, $K$, $r_{\W}$ and $r_{\M}$ which define the model enter in fact only in the combinations $\alpha=r_{\M}/r_{\W}$ and $K_{\E}=K\sqrt{r_{\W}}$. In particular, this explains the behavior of $L$ shown in figure~\ref{shift}.

In order to do so, we recast the dynamics of the model in terms of stochastic differential equations. Let us denote by $i$ (${}=1,2,\ldots,M$) the position of the deme. Then the state of the system is identified by the $2M$-dimensional vector $\vec{x}=(w_{1},m_{1},\ldots,w_{M},m_{M})$. Thus the algorithm described in the previous section can be represented by a master equation of the form
\begin{equation}\label{eq:master}
	\frac{\partial P(\vec{x},t)}{\partial t} = \sum_{A}\left[ t_A( \vec{x} - \vec{r}^A )P( \vec{x} - \vec{r}^A ,t) - t_A( \vec{x})P( \vec{x},t) \right],
\end{equation}
where the index $A$ runs over all the allowed types of events that lead to a change in $\vec{x}$ (birth, death, migration to a neighboring deme, etc.), $t_A$ is the probability of such an event per unit time and $\vec{r}^A$ is the resulting variation of the $\vec{x}$ vector. The expressions of $t_{A}$ and $\vec{r}^{A}$ for each allowed event $A$ are detailed in~table~\ref{table:master}.

Expanding equation~(\ref{eq:master}) to first order in $1/K$ (see, e.g., \cite[chap.~X]{vankampen}) leads to a Fokker-Planck equation, and a corresponding set of Langevin equations can then be found. Under the assumptions that $m_i \approx m_{i+1}$ and $w_i \approx w_{i+1}$, we may approximate the $m_i$ and $w_i$ by the continuous functions $m(x,t)$ and $w(x,t)$. If we further assume that $r_{\W}, r_{\M} \ll 1$, and that stochastic deviations from the average diffusion term are negligible, these equations read:
\begin{eqnarray}
	\frac{\partial w}{\partial t} &=& r_{\W} w(1-m-w) + \frac{\partial^{2}w}{\partial x^{2}} \nonumber \\&&\qquad {}+ \sqrt{2 \frac{w(1-m-w)}{K}}\eta^{\W}(x,t) - \sqrt{2\frac{mw}{K}}\eta^{\W,\M}(x,t); \label{system1}\\
	\frac{\partial m}{\partial t} &=& r_{\M} m(1-m-w) + \frac{\partial^{2} m}{\partial x^{2}}  \nonumber \\ && \qquad  {}+ \sqrt{2 \frac{m(1-m-w)}{K}}\eta^{\M}(x,t)+ \sqrt{2\frac{mw}{K}}\eta^{\W,\M}(x,t).\label{system2}
\end{eqnarray}
In this expression, the Gaussian noises $\eta^{\M}$, $\eta^{\W}$ and $\eta^{\W,\M}$ are uncorrelated, and one has, for instance, $\langle \eta^{\M}(x,t) \eta^{\M}(x',t')\rangle = \delta(x-x')\delta(t-t')$. This set of equations corresponds to a stochastic reaction-diffusion system, where the reaction term is logistic, and where, by construction, the diffusion constant is equal to~1. Notice that the last term corresponds to the stochastic replacement of a mutant by a wildtype individual (or conversely) and is responsible for stochastic fluctuations within a full deme.

The equations can be made nondimensional by setting
\begin{equation}
	X = \sqrt{r_{\W}}x; \qquad T = r_{\W} t; \qquad \alpha = r_{\M}/r_{\W}.
\end{equation}
We obtain therefore
\begin{eqnarray}
	\frac{\partial w}{\partial T} &=& w(1-w-m) + \frac{\partial^{2}w}{\partial X^{2}} \label{ad-system1}\\
	&&\qquad\qquad {}+ \sqrt{2\frac{ w(1-w-m)}{K\sqrt{r_{\W}}}}\eta^{\W}(X,T) - \sqrt{\frac{ 2 wm }{K\sqrt{r_{\W}}}}\eta^{\W,\M}(X,T); \nonumber \\
	\frac{\partial m}{\partial T} &=& \alpha w(1-w-m) + \frac{\partial^{2} m}{\partial X^{2}}\label{ad-system2} \\
	&&\qquad\qquad {} + \sqrt{2 \frac{ m(1-m-w)}{K\sqrt{r_{\W}}}}\eta^{\M}(X,T)  + \sqrt{\frac{ 2 wm }{K\sqrt{r_{\W}}}}\eta^{\W,\M}(X,T). \nonumber	
	\end{eqnarray}
The nondimensionalized equations reveal, as anticipated, that the problem only depends on two relevant parameters: $\alpha=r_{\M}/r_{\W}$ and $K_{\E}=K\sqrt{r_{\W}}$.
\subsection*{Survival probability of a branching process}
\label{sec:surv-prob-branch}
The survival probability of a linear branching process with birth rate $r_\M$ and death rate $1$ can be easily determined by the following discrete reasoning: let us denote the total number of individuals $\sum_i K \times m_i$ by $m_\mathrm{tot} $, and consider the probability $P_{m_\mathrm{tot}}$ that a population of $m_\mathrm{tot}$ individuals will survive. Diffusion events do not change $m_\mathrm{tot}$; it is only affected by duplication events (births or deaths). However, death events are always $(1-r_{\M})$ times less likely than birth events (see the definition of the model). Thus, a given duplication event is a birth with probability $1/(2-r_{\M})$ and a death with probability $(1-r_{\M})/(2-r_{\M})$. By conservation of the probability after such an event we have 
\begin{equation}
	P_{m_\mathrm{tot}} = \frac{1-r_{\M}}{2-r_{\M}}P_{m_\mathrm{tot}-1} + \frac{1}{2-r_{\M}}P_{m_\mathrm{tot}+1},
\end{equation}
with the boundary conditions $P_0 = 0$ and $P_{\infty} = 1$. We obtain therefore 
\begin{equation}
	P_{m_\mathrm{tot}} = 1 - (1-r_{\M})^{m_\mathrm{tot}}.
\end{equation}
Thus the probability that the population stemming from one single mutant will survive is given by
\begin{equation}
 	P_1 = r_{\M}.
\end{equation} 
In the bulk, the only possible events are the replacement of a mutant by a wildtype individual (or the opposite), which take place with the same probability. Thus the size of an isolated mutant population in the bulk undergoes a critical branching process in the presence of an infinite reservoir of wildtype individuals, and its survival probability vanishes.
\subsection*{Heuristic derivation of the differential equation for the surfing probability}
\label{subsec:diffeq-app}
Here, we provide a heuristic rational for the differential equation \eqref{eq:BRW-survival-proba} for the surfing probability $u(x)$. Let us consider the introduction of a mutant at time $\tau$ and position $\xi$. We denote the probability to find a mutant at a position $x$ and at a later time $t$ by $p(x,t|\xi, \tau)$. Now, let us place ourselves in the conditions in which
\begin{itemize}
\item $t$ is close to $\tau$, so that the average mutant population is not yet very large. In this case, the wildtype profile $\langle w \rangle$ is not yet perturbed by the mutants, and in particular, it is steady in the moving frame, i.e., in the frame that goes at the same velocity $v$ as the wildtype wave: $\langle w \rangle(x,t) = \langle w \rangle_\mathrm{init}(x)$, where $\langle w\rangle_\mathrm{init}(x)$ is the initial average profile of the wildtype wave.
\item $u$ is so small that, most of the time, mutants will disappear from the front.
\end{itemize}
In this case, if we find a mutant at $x$ and $t$, its situation is essentially the same as if it had just been introduced in a wave consisting only of wildtype individuals, since $\langle w\rangle(x,t) = \langle w \rangle_\mathrm{init}(x)$ and since, if there are other mutants in the wave at $t$, they will probably not perturb its dynamics. Indeed, for small $u$, other mutants will disappear, in most realizations, before getting a chance to interact effectively with the mutant we consider. Therefore, for this mutant at $x$ and $t$, the probability to fix is by definition $u(x)$.

We may therefore decompose the probability $u(\xi)$ as follows:
\begin{equation}
\label{decompose}
	u(\xi) = \int_{-\infty}^{+\infty} \D x \; u(x)\, p(x,t|\xi, \tau).
\end{equation}
However, this formula is an overestimate of $u(\xi)$. Rare realizations in which two mutants are present at $t$, and in which the issues of both survive, should be counted as one single fixation event, but are in fact double-counted by the formula (\ref{decompose}). Therefore, we expect a negative correction of order $u^2$ when $u$ becomes larger.

If, however, we neglect for the moment this correction, differentiating equation (\ref{decompose}) with respect to $t$ leads to
\begin{equation}
\label{diff}
	0 = \int_{-\infty}^{+\infty} \D x \; u(x)\, \partial_{t} p(x,t|\xi, \tau).
\end{equation}
Notice that, in fact, $p(x,t|\xi, \tau) = \langle m \rangle(x,t)$. Since the mutant population is not very large at $t$, we can neglect the term $-r_m m^2$ in equation (\ref{system2}), and replace $-r_m mw$ by $-r_m m\langle w\rangle_{\mathrm{init}}$. Therefore, in the frame moving with the velocity $v$ of the wave, equation (\ref{system2}) becomes, for $p$:
\begin{equation}
	\label{ODE}
	\frac{\partial p}{\partial t} = r_m\left(1-\langle w \rangle_{\mathrm{init}}\right)p + v\frac{\partial p}{\partial x} + \frac{\partial^2 p}{\partial x^{2}}.
\end{equation}
Upon substituting this expression of  $\partial_t p$ in equation (\ref{diff}), integrating by parts, and noticing that the equation is valid for all $\xi$, we obtain the necessary condition
\begin{equation}
	\label{linear}
	0 = r_m\left(1-\langle w \rangle_\mathrm{init}\right)u - v\frac{\partial u}{\partial x} + \frac{\partial^2 u}{\partial x^{2}}.
\end{equation}

Because of the assumptions that were used in its derivation, this equation is only valid when $u$ is small, i.e., close to the bulk of the wave. However, far ahead of the front, equation (\ref{linear}) does \emph{not} predict the observed saturation of $u$ at $r_m$. We attribute this to the fact that we neglected corrections of order $u^2$. Therefore, we may add a \emph{phenomenological} non-linear term to equation (\ref{linear}):
\begin{equation}
\label{NL-app}
	0 = r_m\left(1-\langle w \rangle_\mathrm{init}\right)u - v\frac{\partial u}{\partial x} + \frac{\partial^2 u}{\partial x^{2}} - u^2.
\end{equation}
This term leaves equation (\ref{linear}) unchanged when $u$ is small, but leads to the correct saturation at $r_m$ far from the front. 
\subsection*{Analysis of the substitution rate}
Our analysis of the substitution rate starts from the observation that the integrand in the expression for the substitution rate in equation \eqref{eq:substitution_rate} has mainly support in the region where $\left<w\right>$ decays exponentially, and $u$ increases exponentially, see Fig.~\ref{curve}. This reflects the tradeoff between high population density (required for the production of mutations) and high surfing probability (required for the fixation of mutations) that determines the substitution process. In the regions that significantly contribute to $G$, we may thus approximate the wild type wave profile by 
\begin{equation}
  \label{eq:w-approx}
  w(x)\approx \exp(- v_\W x /2 ) \;,
\end{equation}
for $x>0$ and $w=1$ otherwise. Here, $v_\W$  is the actual speed observed for the wild type wave. Secondly, we approximate $u$ by 
\begin{equation}
  \label{eq:u-approx}
  u(x) \approx r_\M \exp(v_\M (x-L)/2) \;,
\end{equation}
for $x<L$, and $u=r_\M$ otherwise. Here, $v_\M=2\sqrt{r_\M}$ is the deterministic speed of a mutant wave.

Using these exponential approximations, we can estimate $G$ as
\begin{equation}
  \label{eq:G}
  G\approx K\int_0^L \D x\; u(x) w(x)\approx 2 K r_\M \frac{\exp\left(-v_\W L/2 \right) - \exp\left(-v_\M L/2 \right)}{v_\M- v_\W}
\end{equation}
Equation \eqref{eq:G} is hard to evaluate for general $K$ and selection strength. However, one can derive an asymptotically correct expression for $G$ in the limit of large $K$ for fixed $s\equiv \alpha-1 \ll1$, where the exponential approximation is the leading order description of the wave profile~\cite{Derrida}. In this limit, the equation for the survival probability describes a Fisher wave running in the $-x$ direction with a cutoff far in the tip of the wave, as discussed after Eq.~\eqref{eq:BRW-survival-proba}. The cutoff (due to the net growth rate being proportional to $ 1-\langle w \rangle$ in \eqref{eq:BRW-survival-proba}) has the effect of lowering the wave speed from the deterministic value $v_\M$ to the wildtype value $v_\W$. With the cut-off at position $L$ one obtains an asymptotic wave speed of $\sqrt{v_\M^2-(2\pi/L)^{2}}$~\cite{Derrida}. For this lowered wave speed to equal the wildtype speed $v_\W$, we find
\begin{eqnarray}
  \label{eq:L-asymptotically-1}
  L &\approx &\frac {2\pi}{\sqrt{v_\M^2-v_\W^2}} \\
  &\approx &\frac{2\pi}{v_\W \sqrt{2s}} \;,   \label{eq:L-asymptotically-2}
\end{eqnarray}
where the last equation holds for $s\ll1$. Since we have $v_{\W,\M}\approx 2\sqrt {r_{\W,\M}}$ in the limit $K\to\infty$, we can now express \eqref{eq:G} in terms of our model parameters, obtaining
\begin{eqnarray}
  \label{eq:G-deterministic}
  G \approx 2K\sqrt{r_\W} \frac{\pi}{\sqrt s} \exp\left(-\frac \pi {\sqrt s} \right)\;,
\end{eqnarray}
which holds for small $s\ll 1$. Notice that $G$ is characterized by an essential singularity for $s\to 0$, which causes very small substitution rates for small $s$, indicating that selection is very inefficient at advancing fronts. 

\begin{figure}[!ht]
\centering
\includegraphics[width=0.6\textwidth]{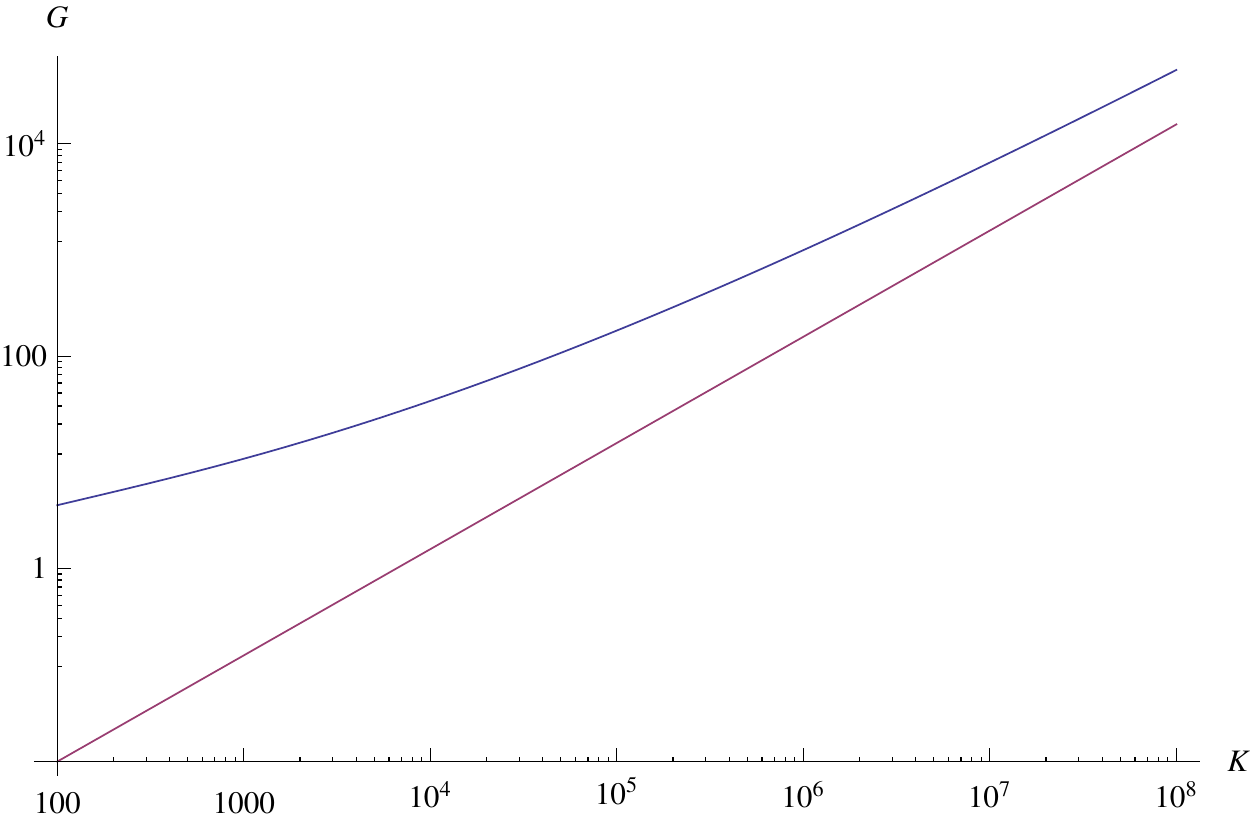}
\caption{{\bf Theoretical approximations for the substitution rate at the front of an advancing population for $r_\W=0.1$, $s=0.1$.} The red curve represents the deterministic approximation \eqref{eq:G-deterministic}, the blue curve corresponds to the approximation in \eqref{eq:G-noisy}, which accounts for the leading order effects of a finite carrying capacity $K$. Notice the large enhancement of the substitution rate due to a finite (even if large) value of $K$.}
\label{fig:G-det-noisy}
\end{figure}

Our analysis neglected so far the effects of a finite carrying capacity $K$. We can account for finite $K$ to leading order by taking advantage of known results for noisy traveling waves, i.e., the fact that to leading order $v_\W$ is given by~\cite{Derrida}
\begin{equation}
  \label{eq:vw}
  v_\W\approx 2 \sqrt{r_\W}\left(1-\frac{\pi^2}{2\ln^2 K_\E}\right) \;.
\end{equation}
Inserting this expression in \eqref{eq:L-asymptotically-1} yields a substitution rate of
\begin{equation}
  \label{eq:G-noisy}
  G\approx \frac{2\pi  K_\E^{1-\pi/\sqrt{s \ln ^2 K_\E+\pi
   ^2}}}{\sqrt{\left(\pi ^2/\ln ^2K_\E\right)+s}}\;.
\end{equation}
In Fig.~\ref{fig:G-det-noisy}, we plot the theoretical predictions for $G$ vs.~$K$, while simulation data are shown in figure~\ref{fig:G/K}. Notice that the expression with finite $K$ stays far below the deterministic limit for any reasonable value of $K$, which results in a non-trivial power law dependence on $K_\E$. From the expression in \eqref{eq:G-noisy}, it is clear that the effect of a finite carrying capacity is important unless $s\gg \pi^2 /\ln^{2}K_\E$, which requires extremely large populations for reasonable selection coefficients. In the opposite quasi-neutral case, the expression for lead $L$ reduces to the position of the cutoff in a noisy Fisher wave, $L\sim \ln K_\E/\sqrt{r_\W}$~\cite{Derrida}.

\section*{Acknowledgments}
This work was done during RL's internship at the Laboratoire Physico-Chimie Curie, UMR 168, Institut Curie, Paris. It was started during a visit by the authors at the Kavli Institute for Theoretical Physics, University of California at Santa Barbara, during the program VIRAL11.  The authors are grateful to the organizers and the host of the program to have made our collaboration possible.

\bibliography{leheHallatschPeliti}

\begin{thebibliography}{10}
\providecommand{\url}[1]{\texttt{#1}}
\providecommand{\urlprefix}{URL }
\expandafter\ifx\csname urlstyle\endcsname\relax
  \providecommand{\doi}[1]{doi:\discretionary{}{}{}#1}\else
  \providecommand{\doi}{doi:\discretionary{}{}{}\begingroup
  \urlstyle{rm}\Url}\fi
\providecommand{\bibAnnoteFile}[1]{%
  \IfFileExists{#1}{\begin{quotation}\noindent\textsc{Key:} #1\\
  \textsc{Annotation:}\ \input{#1}\end{quotation}}{}}
\providecommand{\bibAnnote}[2]{%
  \begin{quotation}\noindent\textsc{Key:} #1\\
  \textsc{Annotation:}\ #2\end{quotation}}
\providecommand{\eprint}[2][]{\url{#2}}

\bibitem{Excoffier:2008hw}
Excoffier L, Ray N (2008) {Surfing during population expansions promotes
  genetic revolutions and structuration}.
\newblock Trends in Ecology {\&} Evolution 23: 347--351.
\bibAnnoteFile{Excoffier:2008hw}

\bibitem{excoffier2009genetic}
Excoffier L, Foll M, Petit R (2009) Genetic consequences of range expansions.
\newblock Annual Review of Ecology, Evolution, and Systematics 40: 481--501.
\bibAnnoteFile{excoffier2009genetic}

\bibitem{fisher}
Fisher RA (1937) The wave of advance of advantageous genes.
\newblock Annals of Eugenics 7: 355--369.
\bibAnnoteFile{fisher}

\bibitem{Hallatschek-theory}
Hallatschek O, Nelson DR (2008) Gene surfing in expanding populations.
\newblock Theoretical Population Biology 73: 158--170.
\bibAnnoteFile{Hallatschek-theory}

\bibitem{edmons}
Edmonds CA, Lillie AS, Cavalli-Sforza LL (2004) Mutations arising in the wave
  front of an expanding population.
\newblock Proceedings of the National Academy of Sciences USA 101: 975--979.
\bibAnnoteFile{edmons}

\bibitem{klopfstein}
Klopfstein S, Currat M, Excoffier L (2006) The fate of mutations surfing on the
  wave of a range expansion.
\newblock Molecular Biology and Evolution 23: 482--490.
\bibAnnoteFile{klopfstein}

\bibitem{hallatschek2007genetic}
Hallatschek O, Hersen P, Ramanathan S, Nelson D (2007) Genetic drift at
  expanding frontiers promotes gene segregation.
\newblock Proceedings of the National Academy of Sciences USA 104:
  19926--19930.
\bibAnnoteFile{hallatschek2007genetic}

\bibitem{hallatschek2010life}
Hallatschek O, Nelson D (2010) Life at the front of an expanding population.
\newblock Evolution 64: 193--206.
\bibAnnoteFile{hallatschek2010life}

\bibitem{travis}
Travis JMJ, M\"unkem\"uller T, Burton OJ, Best A, Dytham C, et~al. (2007)
  Deleterious mutations can surf to high densities on the wave front of an
  expanding population.
\newblock Molecular Biology and Evolution 24: 2334--2343.
\bibAnnoteFile{travis}

\bibitem{miller2010original}
Miller J (2010) Survival of mutations arising during invasions.
\newblock Evolutionary Applications 3: 109--121.
\bibAnnoteFile{miller2010original}

\bibitem{Burton:2008gg}
Burton OJ, Travis JMJ (2008) {Landscape structure and boundary effects
  determine the fate of mutations occurring during range expansions}.
\newblock Heredity 101: 329--340.
\bibAnnoteFile{Burton:2008gg}

\bibitem{munkemuller:2010dw}
M\"{u}nkem\"{u}ller T, Travis MJ, Burton OJ, Schiffers K, Johst K (2011)
  {Density-regulated population dynamics and conditional dispersal alter the
  fate of mutations occurring at the front of an expanding population}.
\newblock Heredity 106: 678--689.
\bibAnnoteFile{munkemuller:2010dw}

\bibitem{shine11}
Shine R, Brown GP, Phillips BL (2011) An evolutionary process that assembles
  phenotypes through space rather than through time.
\newblock Proceedings of the National Academy of Sciences USA 108: 5708--5711.
\bibAnnoteFile{shine11}

\bibitem{lachmuth}
Lachmuth S, Durka W, Schurr FM (2011) Differentiation of reproductive and
  competitive ability in the invaded range of \textsl{Senecio inaequidens}: the
  role of genetic {A}llee effects, adaptive and nonadaptive evolution.
\newblock New Phytologist 192: 529--541.
\bibAnnoteFile{lachmuth}

\bibitem{brown2007invasion}
Brown G, Shilton C, Phillips B, Shine R (2007) Invasion, stress, and spinal
  arthritis in cane toads.
\newblock Proceedings of the National Academy of Sciences USA 104:
  17698--17700.
\bibAnnoteFile{brown2007invasion}

\bibitem{moreau2011deep}
Moreau C, Bh{\'e}rer C, V{\'e}zina H, Jomphe M, Labuda D, et~al. (2011) Deep
  human genealogies reveal a selective advantage to be on an expanding wave
  front.
\newblock Science 334: 1148--1150.
\bibAnnoteFile{moreau2011deep}

\bibitem{van2010gradual}
Van~Bocxlaer I, Loader S, Roelants K, Biju S, Menegon M, et~al. (2010) Gradual
  adaptation toward a range-expansion phenotype initiated the global radiation
  of toads.
\newblock Science 327: 679--682.
\bibAnnoteFile{van2010gradual}

\bibitem{kimura}
Kimura M, Weiss GH (1964) The stepping stone model of population structure and
  the decrease of genetic correlation with distance.
\newblock Genetics 49: 561--576.
\bibAnnoteFile{kimura}

\bibitem{haldane1927mathematical}
Haldane J (1927) A mathematical theory of natural and artificial selection,
  {P}art {V}: {S}election and mutation.
\newblock In: Mathematical Proceedings of the Cambridge Philosophical Society.
  Cambridge University Press, volume~23, pp. 838--844.
\bibAnnoteFile{haldane1927mathematical}

\bibitem{Hallatschek-linear}
Hallatschek O (2011) The noisy edge of traveling waves.
\newblock Proceedings of the National Academy of Sciences USA 108: 1783--1787.
\bibAnnoteFile{Hallatschek-linear}

\bibitem{harris2002theory}
Harris TE (2002) The theory of branching processes.
\newblock New York: Dover Publications.
\newblock Reprint. Originally published by Springer Verlag, 1963.
\bibAnnoteFile{harris2002theory}

\bibitem{Derrida}
Brunet E, Derrida B (1997) Shift in the velocity of a front due to a cutoff.
\newblock Phys Rev E 56: 2597--2604.
\bibAnnoteFile{Derrida}

\bibitem{sniegowski2010beneficial}
Sniegowski P, Gerrish P (2010) Beneficial mutations and the dynamics of
  adaptation in asexual populations.
\newblock Philosophical Transactions of the Royal Society B: Biological
  Sciences 365: 1255--1263.
\bibAnnoteFile{sniegowski2010beneficial}

\bibitem{allee1931co}
Allee W (1931) Co-operation among animals.
\newblock American Journal of Sociology : 386--398.
\bibAnnoteFile{allee1931co}

\bibitem{Murray}
Murray JD (2007) Mathematical Biology: I. An Introduction.
\newblock Berlin: Springer, 3rd edition.
\bibAnnoteFile{Murray}

\bibitem{vankampen}
{van}~Kampen NG (2007) Stochastic Processes in Physics and Chemistry.
\newblock Amsterdam: North Holland, 3rd edition.
\bibAnnoteFile{vankampen}

\end{thebibliography}
\newpage
\section*{Figure Legends}
\begin{description}
\item[Figure 1:] {\bf Fixation probability $u(x)$ as a function of the position $x$ where the mutant is introduced (for $r_{\W} = 0.1$, $r_{\M} = 0.11$ and $K=100$).} The wildtype wave profile (arbitrarily rescaled) is shown by the dashed line. The probability profile $u(x)$ virtually vanishes in the bulk of the wave, but suddenly rises at a characteristic distance $L$ from the front, and then saturates. The shape of the function $u(x)$ is characterized by $L$, defined as the distance between the two points at which the curves reach half of their saturation values, by the characteristic width $L'$ over which the curve rises, and by its saturation height $h$.
\item[Figure 2:] {\bf The measured characteristic distance $L$, rescaled by the factor $r_{\M}/\sqrt{r_{\W}}$, as a function of $\ln(K\sqrt{r_{\W}})$.} The data shown corresponds to various values of $K$, $r_{\W}$ and $r_{\M}$. Values of $K$ range from 100 to 1000, and values of $r_{\W}$ and $r_{\M}$ range from 0.05 to 0.5. The data points corresponding to the same value of $\alpha\equiv r_{\M}/r_{\W}$ group on straight lines, with slopes weakly dependent on $\alpha$. Compare~(\ref{shift-eqn}).
\item[Figure 3:] {\bf Fixation probability profile $u(x)$ for $K=100$, $r_w = 0.1$, and different values of $r_{\M}$.} The dots represent simulation data and the continuous line corresponds to the result of the numerical integration. The dashed line represents the (arbitrarily rescaled) average profile $\langle w \rangle$ of an all wildtype wave.
\item[Figure 4:] {\bf Fixation probability profile $u(x)$ for $r_m = r_w = 0.1$, and different values of $K$.} The dots represent simulation data and the continuous line corresponds to the result of the numerical integration. As before, the dashed line represents $\langle w \rangle_\mathrm{init}$, arbitrarily rescaled.
\item[Figure 5:] {\bf The substitution rate at the front of an advancing population compared to the neutral substitution rate is described by the function $G\equiv K\int\D x\; \left<u(x)\right>\left<w(x)\right>$, which is displayed here as a function of the selective advantage of the mutants.} Notice that the $G$ axis is logarithmic. Blue, red and yellow symbols correspond to the carrying capacity $K=100$, $K=316$ and $K=1000$. All curves approach $1$ in the neutral case, $s=0$, in which substitution and mutation rate equal. Notice the rather slow increase of substitution rates with increasing selection coefficient, for small values of $s$: even for $s=10\%$ and the highest carrying capacity, the substitution rates are merely 4 times higher than the neutral baseline, illustrating the ineffectiveness of selection at expanding fronts. For larger selection coefficients, however, the substitution rate grows roughly exponentially with $s$. 
\item[Figure 6:] {\bf Substitution rate $G$ divided by the deme carrying capacity $K$ as a function of $K$ for several values of the selection coefficient $s=r_{\M}/r_{\W}-1$ of the mutant.} It appears quite clearly that the substitution rate increases less than linearly with $K$, as suggested by equation~(\ref{eq:G-noisy}).
\item[Figure 7:] {\bf Theoretical approximations for the substitution rate at the front of an advancing population for $r_\W=0.1$, $s=0.1$.} The red curve represents the deterministic approximation \eqref{eq:G-deterministic}, the blue curve corresponds to the approximation in \eqref{eq:G-noisy}, which accounts for the leading order effects of a finite carrying capacity $K$. Notice the large enhancement of the substitution rate due to a finite (even if large) value of $K$.
\end{description}

\newpage
\section*{Tables}
\begin{table}[!ht]
\caption{\bf{Transition probabilities}}
	\begin{center}
	\begin{tabular}{|p{4.5 cm}| p{5cm}| p{3cm}|}
	\hline
	Event & Probability per timestep & Change in $\vec{x}$ \\
	\hline
	Birth of a wildtype individual in deme $i$ & $t_A = \frac{1}{M} w_i (1-w_i-m_i)$ & $r^A_{w_i} = +1/K$ \\ 
	Death of a wildtype individual in deme $i$ & $t_A = \frac{(1-r_{\W})}{M} w_i (1-w_i-m_i)$ & $r^A_{w_i} = -1/K$ \\
	Birth of a mutant in deme $i$ & $t_A = \frac{1}{M} m_i (1-m_i-w_i)$ & $r^A_{\M_i} = +1/K$ \\ 
	Death of a mutant in deme $i$ & $t_A = \frac{(1-r_{\M})}{M} m_i (1-w_i-m_i)$ & $r^A_{m_i} = -1/K$ \\
	Replacement of a wildtype individual 
	by a mutant in deme $i$ & $t_A = \frac{1}{M} w_i m_i$ & $r^A_{w_i} = -1/K$, $r^A_{m_i} = +1/K$ \\
	Replacement of a mutant 
	by a wildtype individual in deme $i$ & $t_A = \frac{1}{M} w_i m_i$ & $r^A_{w_i} = +1/K$, $r^A_{m_i} = -1/K$ \\
	\hline
	A wildtype individual from deme $i$ 
	comes to the neighboring deme $j$ & $t_A = \frac{1}{M} w_i (1-w_j-m_j)$ & $r^A_{w_i} = -1/K$, $r^A_{w_j} = +1/K$ \\
	A mutant from deme $i$ 
	comes to the neighboring deme $j$ & $t_A = \frac{1}{M} m_i(1-w_j-m_j)$ & $r^A_{m_i} = -1/K$, $r^A_{m_j} = +1/K$ \\
	A wildtype individual from deme $i$ exchanges place 
	with a mutant from the neighboring deme $j$ & $t_A = \frac{1}{M} m_j w_i$ & $r^A_{w_i} = -1/K$, $r^A_{w_j} = +1/K$, $r^A_{m_i} = +1/K$, $r^A_{m_j} = -1/K$\\
	\hline
	\end{tabular}
	\end{center}
\begin{flushleft}Transition probabilities for the different events $A$ appearing in the master equation~(\ref{eq:master}).
\end{flushleft}
\label{table:master}

\end{table}

\end{document}